\documentclass[aps,pra,10pt,letterpaper,twocolumn,superscriptaddress,showpacs,preprintnumbers]{revtex4}

\usepackage{etoolbox}
\usepackage{ulem}

\usepackage{color}
\usepackage{graphicx}
\usepackage{epstopdf}
\usepackage{bm}
\usepackage{dcolumn}
\usepackage{verbatim}
\usepackage{mathbbol}
\usepackage{dcolumn}
\usepackage{amsmath}
\usepackage{placeins}
\usepackage{mathrsfs}
\usepackage{MnSymbol}
\usepackage[colorlinks,linkcolor=red,citecolor=blue]{hyperref}

\DeclareTextFontCommand{\emph}{\it}

\definecolor{ReviseColorA}{rgb}{0.4,0,0.6}
\definecolor{ReviseColorB}{rgb}{0,0,0.8}
\newtoggle{ReviseToggle}

\definecolor{LightGrey}{rgb}{0.6,0.6,0.6}

\begin{document}

\title{Sub-cycle valleytronics: control of valley polarization using few-cycle linearly polarized pulses}

\newcommand{\mbi}{Max Born Institute, Berlin, Germany}
\newcommand{\uam}{Dpto. F\'isica Te\'orica de la Materia Condensada, Universidad Aut\'onoma de Madrid, Madrid, Spain}
\newcommand{\tu}{Technische Universit\"at Berlin, Berlin, Germany}
\newcommand{\humboldt}{Department of Physics, Humboldt University, Berlin, Germany}
\newcommand{\imperial}{Blackett Laboratory, Imperial College London, London, United Kingdom}

\author{\'Alvaro Jim\'enez-Gal\'an}\email{jimenez@mbi-berlin.de} \affiliation{\mbi}
\author{Rui Silva}\email{ruiefdasilva@gmail.com} \affiliation{\uam}
\author{Olga Smirnova} \affiliation{\mbi}\affiliation{\tu}
\author{Misha Ivanov} \affiliation{\mbi}\affiliation{\humboldt}\affiliation{\imperial}

\date{\today}
\toggletrue{ReviseToggle}

%%
%%   ABSTRACT
%%
\begin{abstract}
So far, selective excitation of a desired valley in the Brillouin zone of a hexagonal two-dimensional material has 
relied on using circularly polarized fields. We theoretically demonstrate a way to induce, control, and read valley polarization in hexagonal 2D materials on a few-femtosecond timescale using a few-cycle, linearly polarized pulse with 
controlled carrier-envelope phase. The valley pseudospin is encoded in the helicity of the emitted high harmonics
of the driving pulse, allowing one to avoid additional probe pulses and permitting one to induce, manipulate 
and read the valley pseudospin all-optically, in one step. High circularity of the harmonic emission 
offers a method to generate highly  elliptic attosecond pulses with a linearly polarized driver, 
in an all-solid-state setup.
\end{abstract}

\maketitle

Generation of few-cycle laser pulses with controlled electric field oscillations under 
the envelope, i.e. controlled carrier envelope phase (CEP), 
catalyzed the development of attosecond physics and technology, providing
tools to control electron dynamics on sub-laser-cycle time scale~\cite{Baltuska:2003aa,Wirth2011}. 
Attosecond technology has now started to develop its potential in condensed matter, in particular via high harmonic generation (HHG) 
~\cite{Ghimire:2010aa, Hohenleutner2015, Vampa2015, Silva:2018aa, Silva:2019aa, Luu:2015aa, Tancogne-Dejean:2017aa}. 
The practical possibility of shaping individual oscillations of an optical laser pulse marks
first steps towards lightwave electronics -- the sub-cycle monitoring and steering of electronic dynamics~\cite{Goulielmakis2007, Krausz2009} in solids, which holds the potential to increase the speed of 
information processing  
from the current GHz to the PHz rate (PHz lightwave electronics.) 

Almost simultaneously with the advent  of CEP-controlled few-cycle pulses, 
the isolation of a single layer of graphene signalled a breakthrough in modern material science and condensed matter physics~\cite{Novoselov666}. 
The fascinating electronic properties of graphene, with carriers following the massless Dirac equation at 
the two points of the Brillouin zone $\mathbf{K}$ and $\mathbf{K}'$, led to a plethora 
of new physical phenomena such as the anomalous integer quantum Hall effect or topological superconductivity~\cite{CastroNeto2009}. 
Similar exotic properties were found also in graphene's insulating counterpart, i.e., hexagonal boron nitride (hBN) and in other 2D materials such as 
transition metal dichalcogenides (TMDs). One of the most promising aspects of these insulating 2D materials is their 
ability to support an extra electronic degree of freedom, i.e., the valley pseudospin, which labels the energy-degenerate extrema of 
the bands and can serve as an additional information carrier for information processing~\cite{Gunawan:2006aa, Xiao:2007aa, Schaibley2016, Vitale:2018aa}.

\begin{figure}%[t]
\begin{center}
\includegraphics[width=\linewidth]{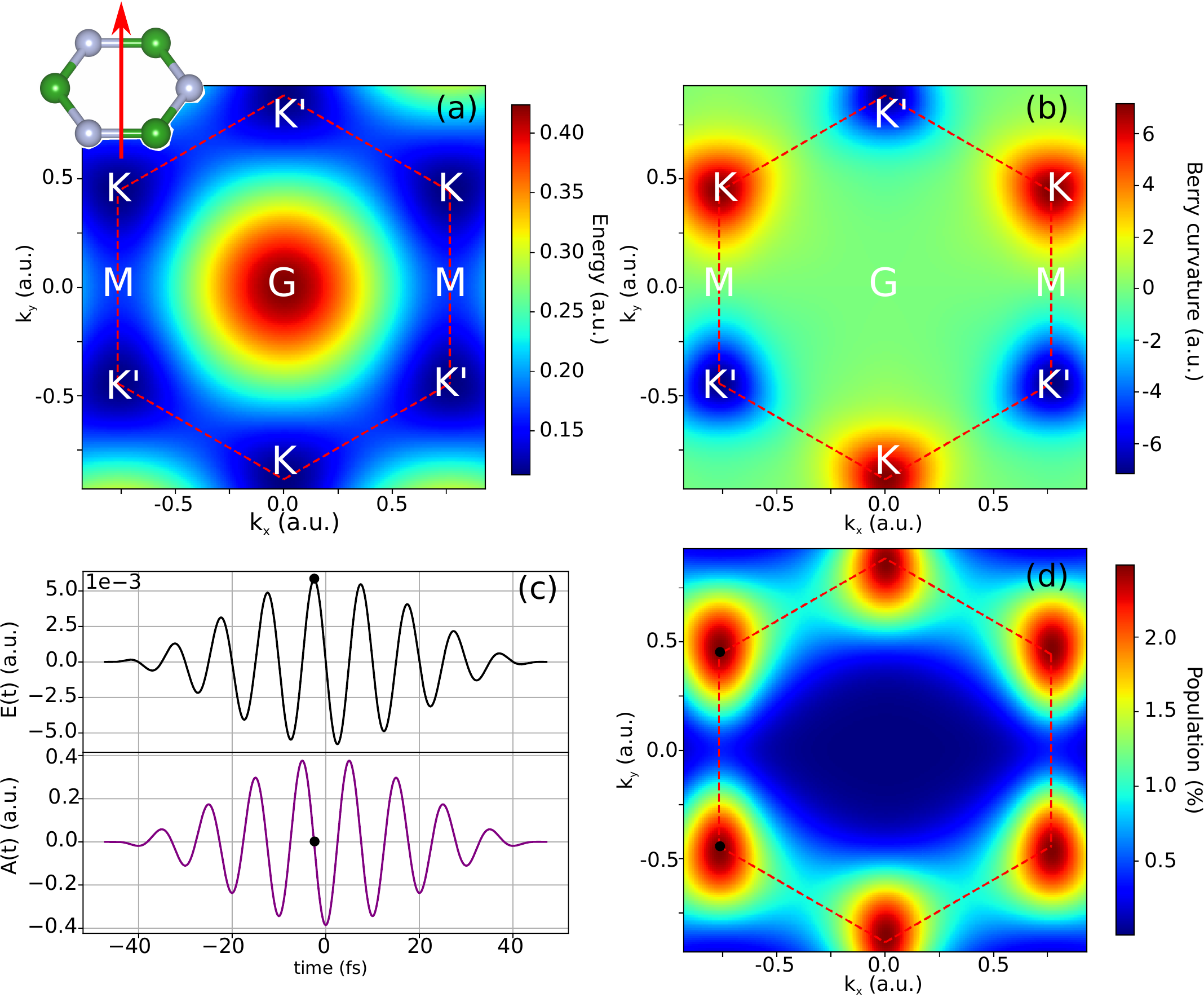}
\caption{\label{fig:bandstructure} Energy (a) and Berry curvature (b) of the first conduction band in hBN. The red dashed lines indicate the FBZ, with the $\mathbf{K}$ and $\mathbf{K}'$ valleys at its vertices forming two triangular sub-lattices.
(c) Electric field (top) and vector potential (bottom), in atomic units, of a 30~fs long pulse carried at $\lambda_L = 3~\mu$m and with field strength $F=0.3$~V/A. The black dots indicate the peak of the electric field, which coincide with the zero of the vector potential.
(d) Electron populations after the long pulse in the first conduction band of hBN.}
\end{center}
\end{figure}

In hexagonal monolayers with broken inversion symmetry such as hBN or TMDs, two different atomic species are 
located at neighbouring sites, creating two inequivalent triangular sub-lattices. This opens a gap at the 
two $\mathbf{K}$ and $\mathbf{K}'$ Dirac points (Fig. \ref{fig:bandstructure}), 
forming energy-degenerate valleys in the landscape of the valence and conduction 
bands~\cite{Gunawan:2006aa, Xiao:2007aa}. 
For a model with one valence and one conduction bands, 
the Bloch electrons carry equal but opposite orbital magnetic 
moment $m(\mathbf{k})$ in the neighbourhood of the two valleys:  $m(\mathbf{k}) = \frac{e}{2\hbar} \Omega(\mathbf{k}) \varepsilon_g(\mathbf{k})$. 
Here $\varepsilon_g (\mathbf{k}) = \varepsilon_c(\mathbf{k}) - \varepsilon_v (\mathbf{k})$ is the 
difference in energy dispersions between the two bands and $\Omega (\mathbf{k})$ is the Berry curvature, which is normal to the monolayer and opposite in both valleys (see Fig.~\ref{fig:bandstructure}a,b). This leads to valley selection rules: right-circularly polarized photons couple to $\mathbf{K}$ while left-circularly 
polarized photons couple to $\mathbf{K}'$~\cite{Xiao:2007aa, Yao:2008aa}. Hence, optical excitation with band-gap-resonant circular drivers generates a valley asymmetry
and 
opens a way to use the valley pseudospin -- the core idea of valleytronics. 
Helicity-induced valley polarization has been demonstrated with circularly-polarized drivers, both in the long-pulse limit~\cite{Mak:2014aa} and for few-cycle pulses~\cite{Oliaei-Motlagh:2019aa}, but its ultrafast control remains one of the main challenges of valleytronics~\cite{Schaibley2016, Vitale:2018aa, Langer:2018aa, Jimenez2019}. 

\begin{figure}[t]
\begin{center}
\includegraphics[width=\linewidth]{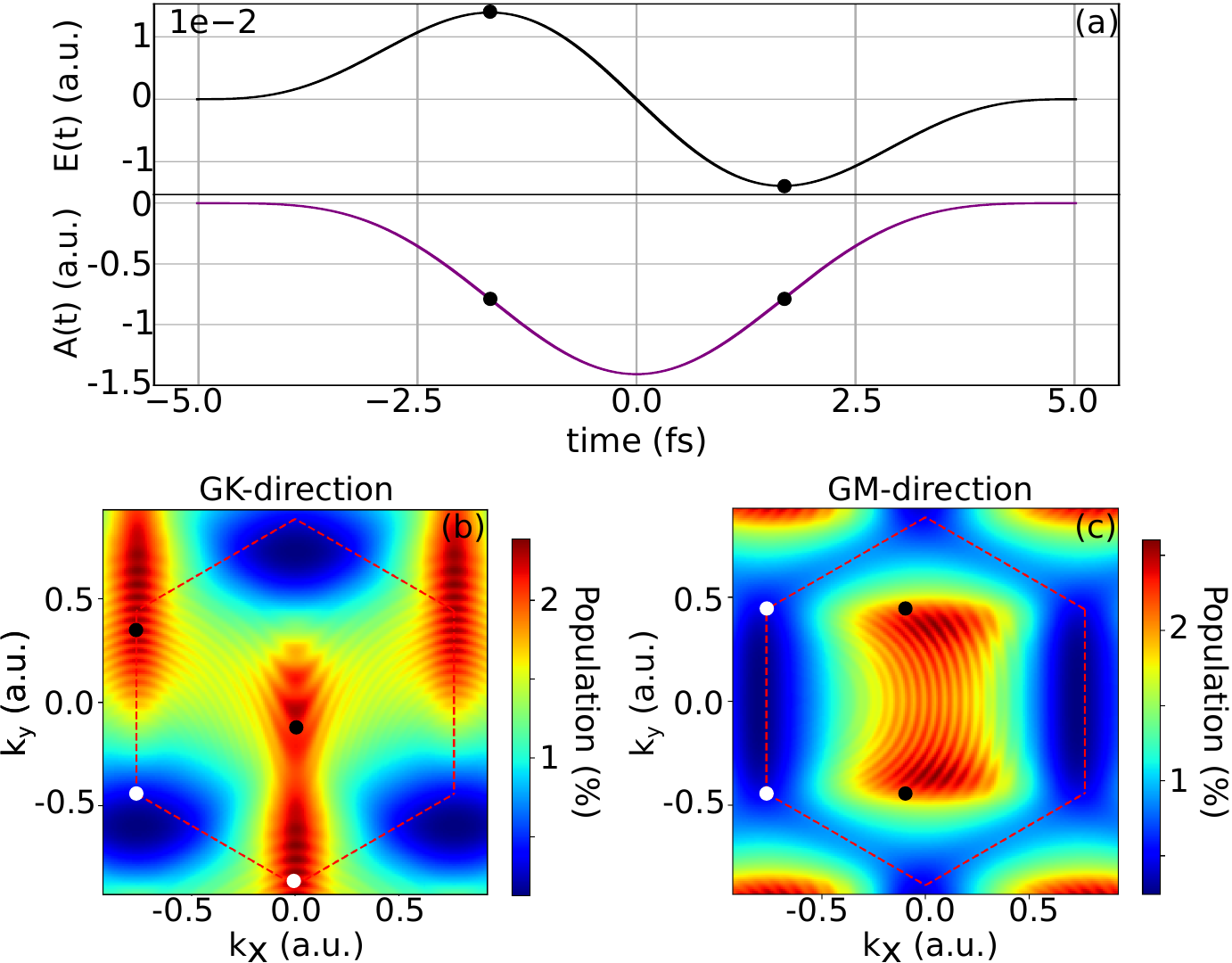}
\caption{\label{fig:single-cycle} (a) Electric field (top) and vector potential (bottom) in a linearly-polarized single-cycle pulse, $\lambda = 3~\mu$m, peak field $1.1$~V/A. The black dots indicate the peaks of the electric field when electron injection occurs. (b,c) Valley populations after the pulse when the pulse polarization is oriented along $\mathbf{G}\mathbf{K}$ and $\mathbf{G}\mathbf{M}$, respectively. White dots show the crystal momenta where non-streaked injection occurs, black dots show where these crystal momenta end after the end of the pulse. }
\end{center}
\end{figure}

The valley selection rules suggest that linearly polarized light couples equally to both valleys, and it is widely accepted that valley polarization with linearly polarized fields is not possible~\cite{Xiao:2007aa, Yao:2008aa, Schaibley2016, Vitale:2018aa, Langer:2018aa}. Here we show that this is not always true. 
We demonstrate that intense, few-cycle, linearly-polarized pulses i) can induce 
high degree of valley polarization, ii) are a tool to achieve sub-cycle control 
of the valley pseudospin, orders of magnitude faster than valley depolarization, 
and iii) offer a probe-free measurement of the valley pseudospin -- the valley asymmetry created by 
the few-cycle pulse is recorded in the helicity of 
harmonics emitted during the interaction. High degree of circularity of the emitted 
harmonics also opens a way to generating highly elliptic attosecond pulses with a solid state setup.

Here we consider hBN, but the concept and physics are general. Figure~\ref{fig:bandstructure}(a,b) shows 
the band structure and the Berry curvature of hBN in its first conduction band, illustrating the energy-minima and the opposite Berry curvatures at the $\mathbf{K}$ and $\mathbf{K}'$ valleys. 
Our numerical method is described in~\cite{Silva2019High,Silva:2019aa,Uzan2020}. Briefly, we first obtain the band structure and the 
transition dipoles from first principles, using density functional theory~\cite{Giannozzi2009}, and then transform into a basis of maximally localized Wannier functions using Wannier90~\cite{Mostofi2008}. For this illustration, we use the two $p_z$ bands of hBN, i.e., the highest valence band and the lowest conduction band. Next, we solve the equation of motion in the density matrix formalism including dephasing as a phenomenological parameter $T_2$, neglecting population relaxation during the short pulse.
After the pulse, we extract the $\mathbf{k}$-resolved electron populations in the first Brillouin zone (FBZ) of the conduction band, shown in Fig.~\ref{fig:bandstructure}(d). This observable can be extracted with angle-resolved photo-emission spectroscopy (ARPES). 

Consider first the effect of a 30 fs, 3~$\mu$m linearly polarized pulse with modest field strength of $F = 0.3$~V/A.
As expected, there is no valley polarization: linearly polarized fields, 
formed by equal superposition of right and left circular drivers, couple equally to both valleys. 

\begin{figure}[t]
\begin{center}
\includegraphics[width=\linewidth]{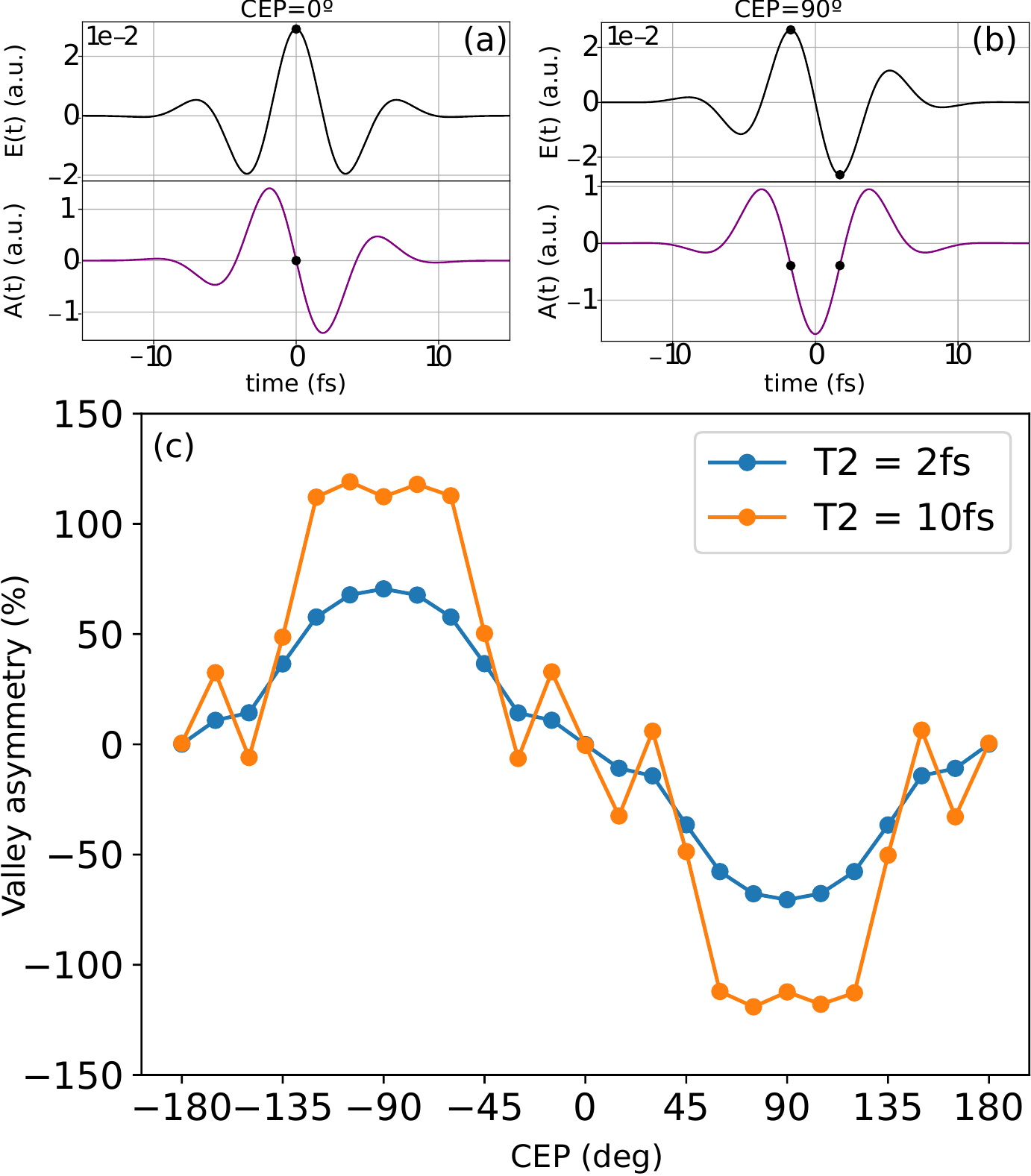}
\caption{\label{fig:cep-control} (a,b) Electric field (top) and vector potential (bottom) of two-cycle pulses,
 $\lambda = 3~\mu$m, peak field $1.5$~V/A, for CEP=0$^\circ$ (a) and CEP=90$^\circ$ (b).  
 (c) CEP control of the valley asymmetry defined by Eq.~\ref{eq:asymmetry}.
}
\end{center}
\end{figure}

Consider now few-cycle pulses, starting with the single-cycle pulse in Fig. \ref{fig:single-cycle}(a),  which allows us to illustrate the relevant physics in the most transparent manner. We shall turn to more realistic pulses in Fig.~\ref{fig:cep-control}.   Figs. \ref{fig:single-cycle}(b,c) show results of our simulations for $\lambda = 3~\mu$m and peak electric field of $F = 1.1$~V/A (still below the damage threshold thanks to the very short pulse duration). 
Strong valley polarization is achieved for the pulse in panel (b),  
no valley polarization is achieved for the pulse in panel (c). 
What is the physics behind this result?

The carrier frequency of the pulse is well below the bandgap of the material. In this regime, electrons are injected into the conduction band near the instantaneous maxima $t_0$ of the electric field, 
marked with black dots in panel (a). Also, the injection occurs near the minima of the bandgap, hence initiating the population in the $\mathbf{K}$ and $\mathbf{K}'$ valleys (white dots in panels (b) and (c)). There is no preference between the two valleys during the injection. Charge injection is followed by light-driven acceleration of electrons and holes inside the bands. The crystal momentum becomes $\mathbf{k}(t) = \mathbf{k}(t_0) + \mathbf{A}_L(t) - \mathbf{A}_L(t_0)$, 
where $\mathbf{A}_L (t)$ is the laser vector potential and the 
crystal momentum at the moment of injection $\mathbf{k}(t_0) $ 
is $\mathbf{K}$ or $\mathbf{K}'$.
After the end of the pulse $\mathbf{A}_L(t)=0$, hence the populations should be located around 
$\mathbf{k}_1(t\to \infty) = \mathbf{K} - \mathbf{A}_L (t_0)$ and 
$\mathbf{k}_2(t\to \infty) = \mathbf{K}' - \mathbf{A}_L (t_0)$.

In long pulses, the maxima of the electric field coincide with the zeroes of the vector potential and $\mathbf{A}_L (t_0) =0$, yielding 
$\mathbf{k}_1(t\to \infty) = \mathbf{K}$ 
and $\mathbf{k}_2(t\to \infty) = \mathbf{K}'$ with equal population, as seen in the long-pulse case of Fig.\ref{fig:bandstructure}.
In few-cycle pulses, however, the vector potential does not necessarily vanish at the peaks of the electric field.  In fact, the two field peaks in Fig. \ref{fig:single-cycle}(a) correspond to {\it the same, 
non-zero, value of the vector potential}. 
In our simple picture,  final electron 
populations should cluster around  $\mathbf{k}_1 = \mathbf{K}-\mathbf{A}_L(t_0)$ and  
$\mathbf{k}_2 = \mathbf{K}'-\mathbf{A}_L(t_0)$ 
(black circles in Fig.~\ref{fig:single-cycle}c) for the field oriented along 
$\mathbf{G}\mathbf{K}$. 
%and $\mathbf{G}\mathbf{M}$, respectively).
If $\mathbf{A}_L(t_0) = \mathbf{K} - \mathbf{G}$, then $\mathbf{k}_1 = 
\mathbf{G}$ and $\mathbf{k}_2 = \mathbf{K}$, leading to high degree of valley polarization, as seen in
Fig.~\ref{fig:single-cycle}(c). 
Due to the symmetry of the lattice, 
valley polarization remains zero when the field is polarized along $\mathbf{G}\mathbf{M}$, as seen in Fig.~\ref{fig:single-cycle}(c).

Results for  more practical few-cycle pulses are shown in Fig.~\ref{fig:cep-control}.
Panels (a,b) show two pulses for the two different CEP values,  CEP=0$^\circ$ (a) and CEP=90$^\circ$ (b). The pulses are carried at $\lambda = 3~\mu$m, with the peak field of $F = 1.5$~V/A. When CEP=0$^\circ$, the electric field has a single dominant peak and the vector potential is zero at this peak, which leads to no valley polarization. As the CEP is varied, the value of the vector potential at the field peaks changes, controlling the valley polarization.
Fig.~\ref{fig:cep-control}(c) shows  CEP control of the valley asymmetry defined as
\begin{equation}\label{eq:asymmetry}
\mathcal{A} = \frac{\rho_\mathbf{K'} - \rho_\mathbf{K}}{(\rho_\mathbf{K'} + \rho_\mathbf{K})/2},
\end{equation}
with $\rho_\mathbf{k}$ the electron population around $\mathbf{k}$.
While the overall dynamics in few-cycle pulses is more complex than in the single-cycle
pulse Fig. \ref{fig:single-cycle}(a) due to the interference of
multiple injection events and multiple Bragg scatterings, 
substantial valley polarization can be achieved and 
controlled by varying the CEP. Such valley control is robust with respect to dephasing. The blue curve in Fig.~\ref{fig:cep-control}(c) shows result for $T_2 = 2$~ fs, well below the period of the field, while the orange curve is obtained for $T_2 = 10$~fs, corresponding to the field period. Both situations show clear valley control. Valley polarization maximises for 
CEP=90$^\circ$, as in Fig. \ref{fig:single-cycle}, and 
vanishes for CEP=0$^\circ$, when the vector potential at the peak of
the field is equal to zero, as expected from the above analysis. To maximize the contrast in panel (c), $\mathbf{A}_L (t_0)$ has to be on the order of the $\mathbf{K}$ - $\mathbf{K}'$ separation.

We now show how the induced valley polarization can be measured all-optically using high harmonic emission. Such emission can be triggered either
by the driving pulse itself, which generates the valley polarization, or 
by a delayed probe pulse, as we have shown in ~\cite{Jimenez2017}. Here we consider the effect of the driving pulse itself and focus on interband electron-hole recombination, which dominates the high-energy part of the harmonic spectrum.

For linear driver, the circularity of the 
harmonic photon emitted during the 
electron-hole recombination is determined by the Berry curvature in 
the emission region, tracking the motion of charges across the Berry curvature 
landscape inside the Brillouin zone. Of course,
such measurement reflects electron dynamics during
the laser pulse,
%(following $\mathbf{k}(t) = \mathbf{k}(t_0) + \mathbf{A}_L(t) - \mathbf{A}_L(t_0)$
%in zero approximation) and is different from 
in contrast to the measurement of valley polarization after the end of the pulse.
% which reflects $\mathbf{k}(t\to \infty) = \mathbf{k}(t_0) - \mathbf{A}_L(t_0)$. 
Nevertheless, it will encode the asymmetry in the 
motion of electrons injected in the  $\mathbf{K}$ or $\mathbf{K}'$   
valleys and their different routes across the landscape of the 
Berry curvature, Fig.~\ref{fig:bandstructure}(b), mapping this difference onto the circularity of the emission.

We begin with CEP=$0^{\circ}$, when 
the pulse has single dominant field maximum (Fig.~\ref{fig:cep-control}(a)). During this maximum, the electrons are injected equally into 
the $\mathbf{K}$  and $\mathbf{K}'$ valleys, with no preference in
the circularity of the recombination harmonics. However, a quarter of
the cycle later, when the vector potential reaches its (negative) maximum, 
electrons from the  $\mathbf{K}$ valley will be driven down in Fig.~\ref{fig:bandstructure}(a,b) into the $\mathbf{K}'-\mathbf{G}$ region, while electrons from the $\mathbf{K}'$ valley will be driven down into the $\mathbf{G}-\mathbf{K}$ region. When electrons are emitted from k-points close to $\mathbf{K}$ ($\mathbf{K}'$), with positive (negative) Berry curvature, the harmonics emitted will have right (left) circular polarization owing to the one-photon valley selection rules. Harmonics emitted close to $\mathbf{G}$, on the other hand, will carry close to linear polarization. Due to the larger band gap, cut-off harmonics are emitted close to $\mathbf{G}$. Whether plateau harmonics are emitted from points closer to $\mathbf{K}$ or $\mathbf{K}'$ is determined, just as for the degree of valley polarization, by the strength of the vector potential and the lattice parameters. In this case, plateau harmonics are mostly emitted at $\mathbf{K}$ (see Supplementary Material). Therefore, we should see highly right-circular harmonics in the energy above the lowest bandgap, with the circularity diminishing towards the highest harmonic energies.

The situation changes for CEP=$90^{\circ}$, when the pulse has 
two dominant  field maxima. Consider the electron motion during the 
quarter cycle after each peak. Thanks to the opposite signs of the 
electric field peaks in these maxima, the asymmetry in the motion of the electrons 
injected during first field peak will be reversed and compensated for 
by the asymmetry in the motion of the electrons injected
during the second peak (see Supplementary Material for numerical confirmation).  Thus, the overall circularity 
of the harmonics should be small for  CEP=$90^{\circ}$.

\begin{figure}[t]
\begin{center}
\includegraphics[width=\linewidth]{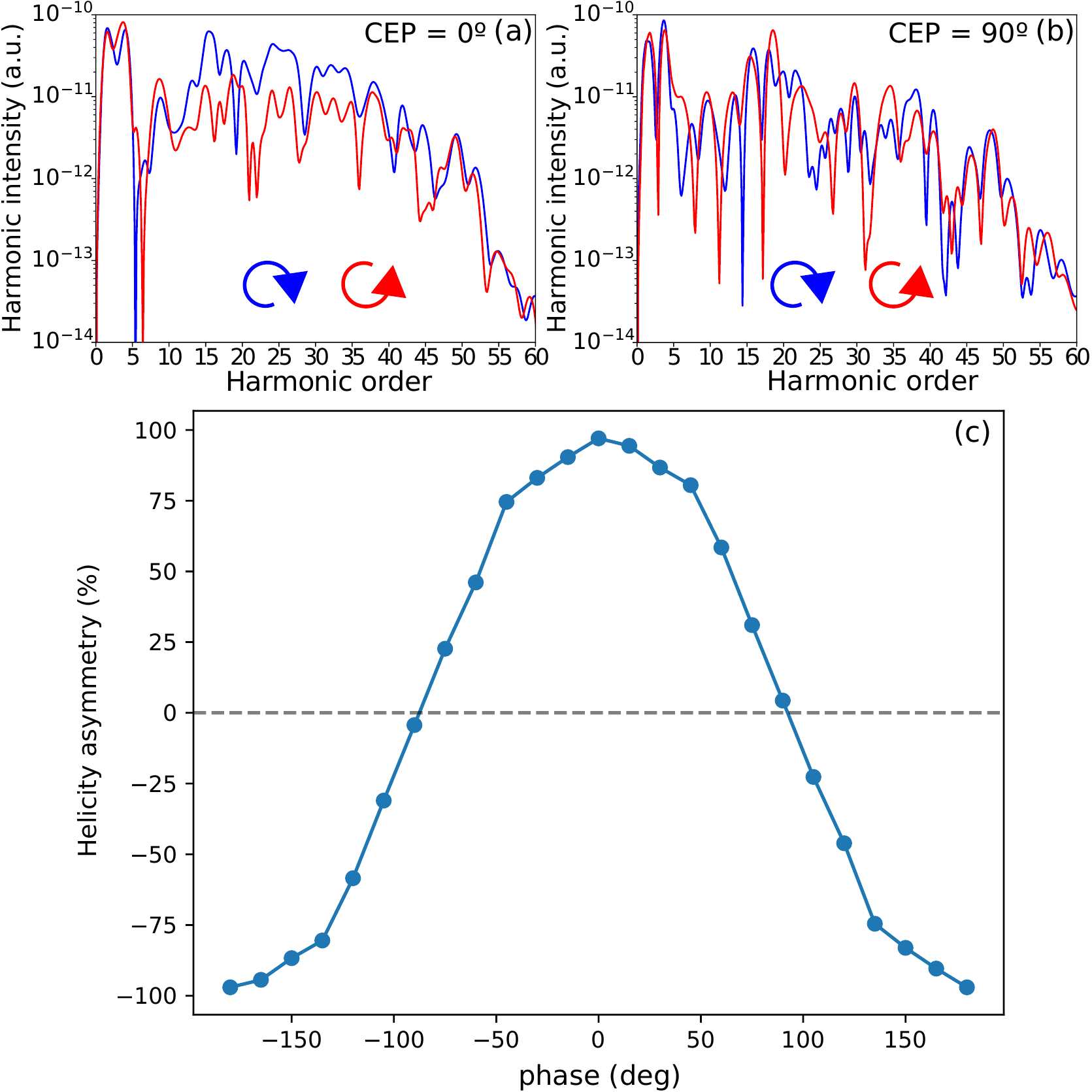}
\caption{\label{fig:hhg} (a,b) Circularity-resolved HHG spectrum for the pulses in Fig.~\ref{fig:cep-control}a,b, respectively. 
(c) Integrated circularity asymmetry as a function of CEP.}
\end{center}
\end{figure}

This is exactly what we find numerically. Figure \ref{fig:hhg}(a,b) shows the harmonic emission resolved
in the right and left circular components, $\hat{e}_\lcirclearrowright = (\hat{e}_x - i\,\hat{e}_y)/\sqrt{2} $, $\hat{e}_\rcirclearrowleft = (\hat{e}_x + i\,\hat{e}_y)/\sqrt{2} $, respectively, for the pulses 
shown in Fig. \ref{fig:cep-control}(a,b), with CEP=0$^\circ$ and CEP=90$^\circ$. 
For CEP=0$^\circ$ all harmonics are highly circular, with the same sign of circularity
determined by the positive Berry curvature at the $\mathbf{K}$ valley.
For CEP=90$^\circ$ the circularity varies widely from negative to positive, reflecting the interference of emission coming from several injection windows. When integrated across all high harmonics above the bandgap, the overall circularity is close to zero.  

To quantify these results, we compute the circularity asymmetry $h = 2(I_{\lcirclearrowright} - I_{\rcirclearrowleft})/(I_\lcirclearrowright + I_\rcirclearrowleft)$, as a function of CEP. 
To obtain the harmonic intensities $I_\lcirclearrowright$ and $I_\rcirclearrowleft$, 
we integrate the energy region between the 20th to 50th harmonic, corresponding to the recombination 
harmonics. The results, presented in Fig.~\ref{fig:hhg}(c), show that  
integrated harmonic polarization changes smoothly from left-circular to linear to right-circular as the CEP varies, crossing zero near CEP$=\pm 90^\circ$.  
Apart from the phase shift, the valley asymmetry after the pulse is faithfully mapped onto the circularity asymmetry 
of the high harmonics emitted during the pulse. This offers an indirect -- but probe-free -- readout of the valley pseudospin. Under experimental conditions, the contribution of long trajectories to the high harmonic spectra is greatly suppressed due to phase matching conditions. To take into account such conditions, we used the dephasing time of $T_2 = 2$~fs~\cite{Floss2018}, similar to that proposed in previous works~\cite{Vampa2014,Hohenleutner2015}. We note, however, that if the contribution of long trajectories is comparable to that of short trajectories, the simple picture above will not hold. In this case, a probe pulse will be necessary, as outlined in ~\cite{Jimenez2017}.

Finally, we note that the high degree of circularity of the HHG spectrum at CEP=0$^\circ$ and 180$^\circ$ will naturally lead to generation of circularly-polarized attosecond pulses. This offers an all-solid-state way of generating such pulses, which are challenging to produce ~\cite{Huang:2018aa, Pisanty:2017aa, Jimenez-Galan2018,Hernandez-Garcia2016,Bandrauk_2017}.

In conclusion, we have demonstrated that few-cycle linearly polarized pulses can induce 
high degree of valley polarization, without violating the 
standard valley selection rules.
Our approach takes advantage of the sub-cycle control of the electron motion by controlling
the oscillations of the electric field under the envelope, and allows for full control of valley polarization
on few-femtosecond time-scale, orders of magnitude faster than the valley polarization lifetime. 
The valley pseudospin can also be measured via polarization-resolved harmonic emission without 
using additional probe pulses. Hence, in one shot and with one single non-resonant pulse, one should be able to write, manipulate and read the valley pseudospin on a few-femtosecond timescale, opening a new way for implementing ultrafast 
valleytronic devices. Just as in atomic systems, the final distribution of electrons in momentum is strongly CEP dependent. For solids, in particular in hexagonal 2D materials, due to the existence of the valley degree of freedom, we are able to take advantage of short pulses to create and control states that are relevant for information processing. Additionally, we have shown that interaction of a nearly single-cycle linearly polarized pulse, stabilized at a CEP=0$^\circ$ or CEP=180$^\circ$, with a gapped graphene system will generate highly circular XUV attosecond pulses.

%%
%%     ACKNOWLEDGMENTS
%%
\begin{acknowledgments}
\'A.J.-G. and M.I. acknowledge support from the Deutsche Forschungsgemeinschaft (DFG) Quantum Dynamics in Tailored Intense Fields (QUTIF) grant IV 152/6-1. R.E.F.S. acknowledges support from the European Research Council Starting Grant (ERC-2016- STG714870). O.S. acknowledges support from the DFG Schwerpunktprogramm 1840 ‘Quantum Dynamics in Tailored Intense Fields’ project SM 292/5-1, and Molecular Electron Dynamics Investigated by Intense Fields and Attosecond Pulses (MEDEA) project, which has received funding from the European Union’s Horizon 2020 research and innovation programme under the Marie Sk\-lodowska-Curie grant agreement no. 641789.
\end{acknowledgments}

%%
%%     BIBLIOGRAPHY
%%
\FloatBarrier
%\bibliography{biblio}

\begin{thebibliography}{34}
\expandafter\ifx\csname natexlab\endcsname\relax\def\natexlab#1{#1}\fi
\expandafter\ifx\csname bibnamefont\endcsname\relax
  \def\bibnamefont#1{#1}\fi
\expandafter\ifx\csname bibfnamefont\endcsname\relax
  \def\bibfnamefont#1{#1}\fi
\expandafter\ifx\csname citenamefont\endcsname\relax
  \def\citenamefont#1{#1}\fi
\expandafter\ifx\csname url\endcsname\relax
  \def\url#1{\texttt{#1}}\fi
\expandafter\ifx\csname urlprefix\endcsname\relax\def\urlprefix{URL }\fi
\providecommand{\bibinfo}[2]{#2}
\providecommand{\eprint}[2][]{\url{#2}}

\bibitem[{\citenamefont{Baltu{\v s}ka et~al.}(2003)\citenamefont{Baltu{\v s}ka,
  Udem, Uiberacker, Hentschel, Goulielmakis, Gohle, Holzwarth, Yakovlev,
  Scrinzi, H{\"a}nsch et~al.}}]{Baltuska:2003aa}
\bibinfo{author}{\bibfnamefont{A.}~\bibnamefont{Baltu{\v s}ka}},
  \bibinfo{author}{\bibfnamefont{T.}~\bibnamefont{Udem}},
  \bibinfo{author}{\bibfnamefont{M.}~\bibnamefont{Uiberacker}},
  \bibinfo{author}{\bibfnamefont{M.}~\bibnamefont{Hentschel}},
  \bibinfo{author}{\bibfnamefont{E.}~\bibnamefont{Goulielmakis}},
  \bibinfo{author}{\bibfnamefont{C.}~\bibnamefont{Gohle}},
  \bibinfo{author}{\bibfnamefont{R.}~\bibnamefont{Holzwarth}},
  \bibinfo{author}{\bibfnamefont{V.~S.} \bibnamefont{Yakovlev}},
  \bibinfo{author}{\bibfnamefont{A.}~\bibnamefont{Scrinzi}},
  \bibinfo{author}{\bibfnamefont{T.~W.} \bibnamefont{H{\"a}nsch}},
  \bibnamefont{et~al.}, \bibinfo{journal}{Nature}
  \textbf{\bibinfo{volume}{421}}, \bibinfo{pages}{611} (\bibinfo{year}{2003}),
  \urlprefix\url{https://doi.org/10.1038/nature01414}.

\bibitem[{\citenamefont{Wirth et~al.}(2011)\citenamefont{Wirth, Hassan,
  Grguras, Gagnon, Moulet, Luu, Pabst, Santra, Alahmed, Azzeer
  et~al.}}]{Wirth2011}
\bibinfo{author}{\bibfnamefont{A.}~\bibnamefont{Wirth}},
  \bibinfo{author}{\bibfnamefont{M.~T.} \bibnamefont{Hassan}},
  \bibinfo{author}{\bibfnamefont{I.}~\bibnamefont{Grguras}},
  \bibinfo{author}{\bibfnamefont{J.}~\bibnamefont{Gagnon}},
  \bibinfo{author}{\bibfnamefont{A.}~\bibnamefont{Moulet}},
  \bibinfo{author}{\bibfnamefont{T.~T.} \bibnamefont{Luu}},
  \bibinfo{author}{\bibfnamefont{S.}~\bibnamefont{Pabst}},
  \bibinfo{author}{\bibfnamefont{R.}~\bibnamefont{Santra}},
  \bibinfo{author}{\bibfnamefont{Z.~a.} \bibnamefont{Alahmed}},
  \bibinfo{author}{\bibfnamefont{a.~M.} \bibnamefont{Azzeer}},
  \bibnamefont{et~al.}, \bibinfo{journal}{Science}
  \textbf{\bibinfo{volume}{334}}, \bibinfo{pages}{195} (\bibinfo{year}{2011}),
  ISSN \bibinfo{issn}{0036-8075}.

\bibitem[{\citenamefont{Ghimire et~al.}(2010)\citenamefont{Ghimire, DiChiara,
  Sistrunk, Agostini, DiMauro, and Reis}}]{Ghimire:2010aa}
\bibinfo{author}{\bibfnamefont{S.}~\bibnamefont{Ghimire}},
  \bibinfo{author}{\bibfnamefont{A.~D.} \bibnamefont{DiChiara}},
  \bibinfo{author}{\bibfnamefont{E.}~\bibnamefont{Sistrunk}},
  \bibinfo{author}{\bibfnamefont{P.}~\bibnamefont{Agostini}},
  \bibinfo{author}{\bibfnamefont{L.~F.} \bibnamefont{DiMauro}},
  \bibnamefont{and} \bibinfo{author}{\bibfnamefont{D.~A.} \bibnamefont{Reis}},
  \bibinfo{journal}{Nature Physics} \textbf{\bibinfo{volume}{7}},
  \bibinfo{pages}{138} (\bibinfo{year}{2010}),
  \urlprefix\url{https://doi.org/10.1038/nphys1847}.

\bibitem[{\citenamefont{Hohenleutner et~al.}(2015)\citenamefont{Hohenleutner,
  Langer, Schubert, Knorr, Huttner, Koch, Kira, and Huber}}]{Hohenleutner2015}
\bibinfo{author}{\bibfnamefont{M.}~\bibnamefont{Hohenleutner}},
  \bibinfo{author}{\bibfnamefont{F.}~\bibnamefont{Langer}},
  \bibinfo{author}{\bibfnamefont{O.}~\bibnamefont{Schubert}},
  \bibinfo{author}{\bibfnamefont{M.}~\bibnamefont{Knorr}},
  \bibinfo{author}{\bibfnamefont{U.}~\bibnamefont{Huttner}},
  \bibinfo{author}{\bibfnamefont{S.~W.} \bibnamefont{Koch}},
  \bibinfo{author}{\bibfnamefont{M.}~\bibnamefont{Kira}}, \bibnamefont{and}
  \bibinfo{author}{\bibfnamefont{R.}~\bibnamefont{Huber}},
  \bibinfo{journal}{Nature} \textbf{\bibinfo{volume}{523}},
  \bibinfo{pages}{572} (\bibinfo{year}{2015}),
  \urlprefix\url{https://doi.org/10.1038/nature14652
  http://10.0.4.14/nature14652}.

\bibitem[{\citenamefont{Vampa et~al.}(2015)\citenamefont{Vampa, Hammond,
  Thir\'e, Schmidt, L\'egar\'e, McDonald, Brabec, Klug, and
  Corkum}}]{Vampa2015}
\bibinfo{author}{\bibfnamefont{G.}~\bibnamefont{Vampa}},
  \bibinfo{author}{\bibfnamefont{T.~J.} \bibnamefont{Hammond}},
  \bibinfo{author}{\bibfnamefont{N.}~\bibnamefont{Thir\'e}},
  \bibinfo{author}{\bibfnamefont{B.~E.} \bibnamefont{Schmidt}},
  \bibinfo{author}{\bibfnamefont{F.}~\bibnamefont{L\'egar\'e}},
  \bibinfo{author}{\bibfnamefont{C.~R.} \bibnamefont{McDonald}},
  \bibinfo{author}{\bibfnamefont{T.}~\bibnamefont{Brabec}},
  \bibinfo{author}{\bibfnamefont{D.~D.} \bibnamefont{Klug}}, \bibnamefont{and}
  \bibinfo{author}{\bibfnamefont{P.~B.} \bibnamefont{Corkum}},
  \bibinfo{journal}{Phys. Rev. Lett.} \textbf{\bibinfo{volume}{115}},
  \bibinfo{pages}{193603} (\bibinfo{year}{2015}),
  \urlprefix\url{https://link.aps.org/doi/10.1103/PhysRevLett.115.193603}.

\bibitem[{\citenamefont{Silva et~al.}(2018)\citenamefont{Silva, Blinov,
  Rubtsov, Smirnova, and Ivanov}}]{Silva:2018aa}
\bibinfo{author}{\bibfnamefont{R.~E.~F.} \bibnamefont{Silva}},
  \bibinfo{author}{\bibfnamefont{I.~V.} \bibnamefont{Blinov}},
  \bibinfo{author}{\bibfnamefont{A.~N.} \bibnamefont{Rubtsov}},
  \bibinfo{author}{\bibfnamefont{O.}~\bibnamefont{Smirnova}}, \bibnamefont{and}
  \bibinfo{author}{\bibfnamefont{M.}~\bibnamefont{Ivanov}},
  \bibinfo{journal}{Nature Photonics} \textbf{\bibinfo{volume}{12}},
  \bibinfo{pages}{266} (\bibinfo{year}{2018}),
  \urlprefix\url{https://doi.org/10.1038/s41566-018-0129-0}.

\bibitem[{\citenamefont{Silva et~al.}(2019{\natexlab{a}})\citenamefont{Silva,
  Jim{\'e}nez-Gal{\'a}n, Amorim, Smirnova, and Ivanov}}]{Silva:2019aa}
\bibinfo{author}{\bibfnamefont{R.~E.~F.} \bibnamefont{Silva}},
  \bibinfo{author}{\bibfnamefont{{\'A}.}~\bibnamefont{Jim{\'e}nez-Gal{\'a}n}},
  \bibinfo{author}{\bibfnamefont{B.}~\bibnamefont{Amorim}},
  \bibinfo{author}{\bibfnamefont{O.}~\bibnamefont{Smirnova}}, \bibnamefont{and}
  \bibinfo{author}{\bibfnamefont{M.}~\bibnamefont{Ivanov}},
  \bibinfo{journal}{Nature Photonics} \textbf{\bibinfo{volume}{13}},
  \bibinfo{pages}{849} (\bibinfo{year}{2019}{\natexlab{a}}),
  \urlprefix\url{https://doi.org/10.1038/s41566-019-0516-1}.

\bibitem[{\citenamefont{Luu et~al.}(2015)\citenamefont{Luu, Garg, Kruchinin,
  Moulet, Hassan, and Goulielmakis}}]{Luu:2015aa}
\bibinfo{author}{\bibfnamefont{T.~T.} \bibnamefont{Luu}},
  \bibinfo{author}{\bibfnamefont{M.}~\bibnamefont{Garg}},
  \bibinfo{author}{\bibfnamefont{S.~Y.} \bibnamefont{Kruchinin}},
  \bibinfo{author}{\bibfnamefont{A.}~\bibnamefont{Moulet}},
  \bibinfo{author}{\bibfnamefont{M.~T.} \bibnamefont{Hassan}},
  \bibnamefont{and}
  \bibinfo{author}{\bibfnamefont{E.}~\bibnamefont{Goulielmakis}},
  \bibinfo{journal}{Nature} \textbf{\bibinfo{volume}{521}},
  \bibinfo{pages}{498} (\bibinfo{year}{2015}),
  \urlprefix\url{https://doi.org/10.1038/nature14456}.

\bibitem[{\citenamefont{Tancogne-Dejean
  et~al.}(2017)\citenamefont{Tancogne-Dejean, M{\"u}cke, K{\"a}rtner, and
  Rubio}}]{Tancogne-Dejean:2017aa}
\bibinfo{author}{\bibfnamefont{N.}~\bibnamefont{Tancogne-Dejean}},
  \bibinfo{author}{\bibfnamefont{O.~D.} \bibnamefont{M{\"u}cke}},
  \bibinfo{author}{\bibfnamefont{F.~X.} \bibnamefont{K{\"a}rtner}},
  \bibnamefont{and} \bibinfo{author}{\bibfnamefont{A.}~\bibnamefont{Rubio}},
  \bibinfo{journal}{Physical Review Letters} \textbf{\bibinfo{volume}{118}},
  \bibinfo{pages}{087403} (\bibinfo{year}{2017}),
  \urlprefix\url{https://link.aps.org/doi/10.1103/PhysRevLett.118.087403}.

\bibitem[{\citenamefont{Goulielmakis et~al.}(2007)\citenamefont{Goulielmakis,
  Yakovlev, Cavalieri, Uiberacker, Pervak, Apolonski, Kienberger, Kleineberg,
  and Krausz}}]{Goulielmakis2007}
\bibinfo{author}{\bibfnamefont{E.}~\bibnamefont{Goulielmakis}},
  \bibinfo{author}{\bibfnamefont{V.~S.} \bibnamefont{Yakovlev}},
  \bibinfo{author}{\bibfnamefont{a.~L.} \bibnamefont{Cavalieri}},
  \bibinfo{author}{\bibfnamefont{M.}~\bibnamefont{Uiberacker}},
  \bibinfo{author}{\bibfnamefont{V.}~\bibnamefont{Pervak}},
  \bibinfo{author}{\bibfnamefont{A.}~\bibnamefont{Apolonski}},
  \bibinfo{author}{\bibfnamefont{R.}~\bibnamefont{Kienberger}},
  \bibinfo{author}{\bibfnamefont{U.}~\bibnamefont{Kleineberg}},
  \bibnamefont{and} \bibinfo{author}{\bibfnamefont{F.}~\bibnamefont{Krausz}},
  \bibinfo{journal}{Science} \textbf{\bibinfo{volume}{317}},
  \bibinfo{pages}{769} (\bibinfo{year}{2007}), ISSN \bibinfo{issn}{0036-8075}.

\bibitem[{\citenamefont{Krausz and Ivanov}(2009)}]{Krausz2009}
\bibinfo{author}{\bibfnamefont{F.}~\bibnamefont{Krausz}} \bibnamefont{and}
  \bibinfo{author}{\bibfnamefont{M.}~\bibnamefont{Ivanov}},
  \bibinfo{journal}{Rev. Mod. Phys.} \textbf{\bibinfo{volume}{81}},
  \bibinfo{pages}{163} (\bibinfo{year}{2009}), ISSN \bibinfo{issn}{0034-6861},
  \urlprefix\url{http://link.aps.org/doi/10.1103/RevModPhys.81.163}.

\bibitem[{\citenamefont{Novoselov et~al.}(2004)\citenamefont{Novoselov, Geim,
  Morozov, Jiang, Zhang, Dubonos, Grigorieva, and Firsov}}]{Novoselov666}
\bibinfo{author}{\bibfnamefont{K.~S.} \bibnamefont{Novoselov}},
  \bibinfo{author}{\bibfnamefont{A.~K.} \bibnamefont{Geim}},
  \bibinfo{author}{\bibfnamefont{S.~V.} \bibnamefont{Morozov}},
  \bibinfo{author}{\bibfnamefont{D.}~\bibnamefont{Jiang}},
  \bibinfo{author}{\bibfnamefont{Y.}~\bibnamefont{Zhang}},
  \bibinfo{author}{\bibfnamefont{S.~V.} \bibnamefont{Dubonos}},
  \bibinfo{author}{\bibfnamefont{I.~V.} \bibnamefont{Grigorieva}},
  \bibnamefont{and} \bibinfo{author}{\bibfnamefont{A.~A.}
  \bibnamefont{Firsov}}, \bibinfo{journal}{Science}
  \textbf{\bibinfo{volume}{306}}, \bibinfo{pages}{666} (\bibinfo{year}{2004}),
  ISSN \bibinfo{issn}{0036-8075},
  \eprint{https://science.sciencemag.org/content/306/5696/666.full.pdf},
  \urlprefix\url{https://science.sciencemag.org/content/306/5696/666}.

\bibitem[{\citenamefont{{Castro Neto} et~al.}(2009)\citenamefont{{Castro Neto},
  Guinea, Peres, Novoselov, and Geim}}]{CastroNeto2009}
\bibinfo{author}{\bibfnamefont{A.~H.} \bibnamefont{{Castro Neto}}},
  \bibinfo{author}{\bibfnamefont{F.}~\bibnamefont{Guinea}},
  \bibinfo{author}{\bibfnamefont{N.~M.~R.} \bibnamefont{Peres}},
  \bibinfo{author}{\bibfnamefont{K.~S.} \bibnamefont{Novoselov}},
  \bibnamefont{and} \bibinfo{author}{\bibfnamefont{A.~K.} \bibnamefont{Geim}},
  \bibinfo{journal}{Rev. Mod. Phys.} \textbf{\bibinfo{volume}{81}},
  \bibinfo{pages}{109} (\bibinfo{year}{2009}),
  \urlprefix\url{https://link.aps.org/doi/10.1103/RevModPhys.81.109}.

\bibitem[{\citenamefont{Gunawan et~al.}(2006)\citenamefont{Gunawan, Shkolnikov,
  Vakili, Gokmen, De~Poortere, and Shayegan}}]{Gunawan:2006aa}
\bibinfo{author}{\bibfnamefont{O.}~\bibnamefont{Gunawan}},
  \bibinfo{author}{\bibfnamefont{Y.~P.} \bibnamefont{Shkolnikov}},
  \bibinfo{author}{\bibfnamefont{K.}~\bibnamefont{Vakili}},
  \bibinfo{author}{\bibfnamefont{T.}~\bibnamefont{Gokmen}},
  \bibinfo{author}{\bibfnamefont{E.~P.} \bibnamefont{De~Poortere}},
  \bibnamefont{and} \bibinfo{author}{\bibfnamefont{M.}~\bibnamefont{Shayegan}},
  \bibinfo{journal}{Physical Review Letters} \textbf{\bibinfo{volume}{97}},
  \bibinfo{pages}{186404} (\bibinfo{year}{2006}),
  \urlprefix\url{https://link.aps.org/doi/10.1103/PhysRevLett.97.186404}.

\bibitem[{\citenamefont{Xiao et~al.}(2007)\citenamefont{Xiao, Yao, and
  Niu}}]{Xiao:2007aa}
\bibinfo{author}{\bibfnamefont{D.}~\bibnamefont{Xiao}},
  \bibinfo{author}{\bibfnamefont{W.}~\bibnamefont{Yao}}, \bibnamefont{and}
  \bibinfo{author}{\bibfnamefont{Q.}~\bibnamefont{Niu}},
  \bibinfo{journal}{Physical Review Letters} \textbf{\bibinfo{volume}{99}},
  \bibinfo{pages}{236809} (\bibinfo{year}{2007}),
  \urlprefix\url{https://link.aps.org/doi/10.1103/PhysRevLett.99.236809}.

\bibitem[{\citenamefont{Schaibley et~al.}(2016)\citenamefont{Schaibley, Yu,
  Clark, Rivera, Ross, Seyler, Yao, and Xu}}]{Schaibley2016}
\bibinfo{author}{\bibfnamefont{J.~R.} \bibnamefont{Schaibley}},
  \bibinfo{author}{\bibfnamefont{H.}~\bibnamefont{Yu}},
  \bibinfo{author}{\bibfnamefont{G.}~\bibnamefont{Clark}},
  \bibinfo{author}{\bibfnamefont{P.}~\bibnamefont{Rivera}},
  \bibinfo{author}{\bibfnamefont{J.~S.} \bibnamefont{Ross}},
  \bibinfo{author}{\bibfnamefont{K.~L.} \bibnamefont{Seyler}},
  \bibinfo{author}{\bibfnamefont{W.}~\bibnamefont{Yao}}, \bibnamefont{and}
  \bibinfo{author}{\bibfnamefont{X.}~\bibnamefont{Xu}},
  \bibinfo{journal}{Nature Reviews Materials} \textbf{\bibinfo{volume}{1}},
  \bibinfo{pages}{16055} (\bibinfo{year}{2016}),
  \urlprefix\url{http://dx.doi.org/10.1038/natrevmats.2016.55}.

\bibitem[{\citenamefont{Vitale et~al.}(2018)\citenamefont{Vitale, Nezich,
  Varghese, Kim, Gedik, Jarillo-Herrero, Xiao, and Rothschild}}]{Vitale:2018aa}
\bibinfo{author}{\bibfnamefont{S.~A.} \bibnamefont{Vitale}},
  \bibinfo{author}{\bibfnamefont{D.}~\bibnamefont{Nezich}},
  \bibinfo{author}{\bibfnamefont{J.~O.} \bibnamefont{Varghese}},
  \bibinfo{author}{\bibfnamefont{P.}~\bibnamefont{Kim}},
  \bibinfo{author}{\bibfnamefont{N.}~\bibnamefont{Gedik}},
  \bibinfo{author}{\bibfnamefont{P.}~\bibnamefont{Jarillo-Herrero}},
  \bibinfo{author}{\bibfnamefont{D.}~\bibnamefont{Xiao}}, \bibnamefont{and}
  \bibinfo{author}{\bibfnamefont{M.}~\bibnamefont{Rothschild}},
  \bibinfo{journal}{Small} \textbf{\bibinfo{volume}{14}},
  \bibinfo{pages}{1801483} (\bibinfo{year}{2018}),
  \urlprefix\url{https://doi.org/10.1002/smll.201801483}.

\bibitem[{\citenamefont{Yao et~al.}(2008)\citenamefont{Yao, Xiao, and
  Niu}}]{Yao:2008aa}
\bibinfo{author}{\bibfnamefont{W.}~\bibnamefont{Yao}},
  \bibinfo{author}{\bibfnamefont{D.}~\bibnamefont{Xiao}}, \bibnamefont{and}
  \bibinfo{author}{\bibfnamefont{Q.}~\bibnamefont{Niu}},
  \bibinfo{journal}{Physical Review B} \textbf{\bibinfo{volume}{77}},
  \bibinfo{pages}{235406} (\bibinfo{year}{2008}),
  \urlprefix\url{https://link.aps.org/doi/10.1103/PhysRevB.77.235406}.

\bibitem[{\citenamefont{Mak et~al.}(2014)\citenamefont{Mak, McGill, Park, and
  McEuen}}]{Mak:2014aa}
\bibinfo{author}{\bibfnamefont{K.~F.} \bibnamefont{Mak}},
  \bibinfo{author}{\bibfnamefont{K.~L.} \bibnamefont{McGill}},
  \bibinfo{author}{\bibfnamefont{J.}~\bibnamefont{Park}}, \bibnamefont{and}
  \bibinfo{author}{\bibfnamefont{P.~L.} \bibnamefont{McEuen}},
  \bibinfo{journal}{Science} \textbf{\bibinfo{volume}{344}},
  \bibinfo{pages}{1489} (\bibinfo{year}{2014}),
  \urlprefix\url{http://science.sciencemag.org/content/344/6191/1489.abstract}.

\bibitem[{\citenamefont{Oliaei~Motlagh
  et~al.}(2019)\citenamefont{Oliaei~Motlagh, Nematollahi, Apalkov, and
  Stockman}}]{Oliaei-Motlagh:2019aa}
\bibinfo{author}{\bibfnamefont{S.~A.} \bibnamefont{Oliaei~Motlagh}},
  \bibinfo{author}{\bibfnamefont{F.}~\bibnamefont{Nematollahi}},
  \bibinfo{author}{\bibfnamefont{V.}~\bibnamefont{Apalkov}}, \bibnamefont{and}
  \bibinfo{author}{\bibfnamefont{M.~I.} \bibnamefont{Stockman}},
  \bibinfo{journal}{Physical Review B} \textbf{\bibinfo{volume}{100}},
  \bibinfo{pages}{115431} (\bibinfo{year}{2019}),
  \urlprefix\url{https://link.aps.org/doi/10.1103/PhysRevB.100.115431}.

\bibitem[{\citenamefont{Langer et~al.}(2018)\citenamefont{Langer, Schmid,
  Schlauderer, Gmitra, Fabian, Nagler, Sch{\"{u}}ller, Korn, Hawkins, Steiner
  et~al.}}]{Langer:2018aa}
\bibinfo{author}{\bibfnamefont{F.}~\bibnamefont{Langer}},
  \bibinfo{author}{\bibfnamefont{C.~P.} \bibnamefont{Schmid}},
  \bibinfo{author}{\bibfnamefont{S.}~\bibnamefont{Schlauderer}},
  \bibinfo{author}{\bibfnamefont{M.}~\bibnamefont{Gmitra}},
  \bibinfo{author}{\bibfnamefont{J.}~\bibnamefont{Fabian}},
  \bibinfo{author}{\bibfnamefont{P.}~\bibnamefont{Nagler}},
  \bibinfo{author}{\bibfnamefont{C.}~\bibnamefont{Sch{\"{u}}ller}},
  \bibinfo{author}{\bibfnamefont{T.}~\bibnamefont{Korn}},
  \bibinfo{author}{\bibfnamefont{P.~G.} \bibnamefont{Hawkins}},
  \bibinfo{author}{\bibfnamefont{J.~T.} \bibnamefont{Steiner}},
  \bibnamefont{et~al.}, \bibinfo{journal}{Nature}
  \textbf{\bibinfo{volume}{557}}, \bibinfo{pages}{76} (\bibinfo{year}{2018}),
  \urlprefix\url{https://doi.org/10.1038/s41586-018-0013-6}.

\bibitem[{\citenamefont{Jim{\'{e}}nez-Gal{\'{a}}n
  et~al.}(2019)\citenamefont{Jim{\'{e}}nez-Gal{\'{a}}n, Silva, Smirnova, and
  Ivanov}}]{Jimenez2019}
\bibinfo{author}{\bibfnamefont{{\'{A}}.}~\bibnamefont{Jim{\'{e}}nez-Gal{\'{a}}n}},
  \bibinfo{author}{\bibfnamefont{R.~E.~F.} \bibnamefont{Silva}},
  \bibinfo{author}{\bibfnamefont{O.}~\bibnamefont{Smirnova}}, \bibnamefont{and}
  \bibinfo{author}{\bibfnamefont{M.~Y.} \bibnamefont{Ivanov}},
  \bibinfo{journal}{arXiv:1910.07398}  (\bibinfo{year}{2019}).

\bibitem[{\citenamefont{Silva et~al.}(2019{\natexlab{b}})\citenamefont{Silva,
  Mart{\'\i}n, and Ivanov}}]{Silva2019High}
\bibinfo{author}{\bibfnamefont{R.~E.~F.} \bibnamefont{Silva}},
  \bibinfo{author}{\bibfnamefont{F.}~\bibnamefont{Mart{\'\i}n}},
  \bibnamefont{and} \bibinfo{author}{\bibfnamefont{M.}~\bibnamefont{Ivanov}},
  \bibinfo{journal}{arXiv:1904.00283}  (\bibinfo{year}{2019}{\natexlab{b}}).

\bibitem[{\citenamefont{Uzan et~al.}(2020)\citenamefont{Uzan, Orenstein,
  Jim{\'{e}}nez-Gal{\'{a}}n, McDonald, Silva, Bruner, Klimkin, Blanchet,
  Arusi-Parpar, Kr{\"{u}}ger et~al.}}]{Uzan2020}
\bibinfo{author}{\bibfnamefont{A.~J.} \bibnamefont{Uzan}},
  \bibinfo{author}{\bibfnamefont{G.}~\bibnamefont{Orenstein}},
  \bibinfo{author}{\bibfnamefont{{\'{A}}.}~\bibnamefont{Jim{\'{e}}nez-Gal{\'{a}}n}},
  \bibinfo{author}{\bibfnamefont{C.}~\bibnamefont{McDonald}},
  \bibinfo{author}{\bibfnamefont{R.~E.~F.} \bibnamefont{Silva}},
  \bibinfo{author}{\bibfnamefont{B.~D.} \bibnamefont{Bruner}},
  \bibinfo{author}{\bibfnamefont{N.~D.} \bibnamefont{Klimkin}},
  \bibinfo{author}{\bibfnamefont{V.}~\bibnamefont{Blanchet}},
  \bibinfo{author}{\bibfnamefont{T.}~\bibnamefont{Arusi-Parpar}},
  \bibinfo{author}{\bibfnamefont{M.}~\bibnamefont{Kr{\"{u}}ger}},
  \bibnamefont{et~al.}, \bibinfo{journal}{Nature Photonics}
  \textbf{\bibinfo{volume}{14}}, \bibinfo{pages}{183} (\bibinfo{year}{2020}),
  ISSN \bibinfo{issn}{1749-4893},
  \urlprefix\url{https://doi.org/10.1038/s41566-019-0574-4}.

\bibitem[{\citenamefont{Giannozzi et~al.}(2009)\citenamefont{Giannozzi, Baroni,
  Bonini, Calandra, Car, Cavazzoni, Ceresoli, Chiarotti, Cococcioni, Dabo
  et~al.}}]{Giannozzi2009}
\bibinfo{author}{\bibfnamefont{P.}~\bibnamefont{Giannozzi}},
  \bibinfo{author}{\bibfnamefont{S.}~\bibnamefont{Baroni}},
  \bibinfo{author}{\bibfnamefont{N.}~\bibnamefont{Bonini}},
  \bibinfo{author}{\bibfnamefont{M.}~\bibnamefont{Calandra}},
  \bibinfo{author}{\bibfnamefont{R.}~\bibnamefont{Car}},
  \bibinfo{author}{\bibfnamefont{C.}~\bibnamefont{Cavazzoni}},
  \bibinfo{author}{\bibfnamefont{D.}~\bibnamefont{Ceresoli}},
  \bibinfo{author}{\bibfnamefont{G.~L.} \bibnamefont{Chiarotti}},
  \bibinfo{author}{\bibfnamefont{M.}~\bibnamefont{Cococcioni}},
  \bibinfo{author}{\bibfnamefont{I.}~\bibnamefont{Dabo}}, \bibnamefont{et~al.},
  \bibinfo{journal}{Journal of Physics: Condensed Matter}
  \textbf{\bibinfo{volume}{21}}, \bibinfo{pages}{395502}
  (\bibinfo{year}{2009}),
  \urlprefix\url{http://stacks.iop.org/0953-8984/21/i=39/a=395502}.

\bibitem[{\citenamefont{Mostofi et~al.}(2008)\citenamefont{Mostofi, Yates, Lee,
  Souza, Vanderbilt, and Marzari}}]{Mostofi2008}
\bibinfo{author}{\bibfnamefont{A.~A.} \bibnamefont{Mostofi}},
  \bibinfo{author}{\bibfnamefont{J.~R.} \bibnamefont{Yates}},
  \bibinfo{author}{\bibfnamefont{Y.-S.} \bibnamefont{Lee}},
  \bibinfo{author}{\bibfnamefont{I.}~\bibnamefont{Souza}},
  \bibinfo{author}{\bibfnamefont{D.}~\bibnamefont{Vanderbilt}},
  \bibnamefont{and} \bibinfo{author}{\bibfnamefont{N.}~\bibnamefont{Marzari}},
  \bibinfo{journal}{Computer Physics Communications}
  \textbf{\bibinfo{volume}{178}}, \bibinfo{pages}{685} (\bibinfo{year}{2008}),
  ISSN \bibinfo{issn}{0010-4655},
  \urlprefix\url{https://www.sciencedirect.com/science/article/pii/S0010465507004936}.

\bibitem[{\citenamefont{Jim{\'{e}}nez-Gal{\'{a}}n
  et~al.}(2017)\citenamefont{Jim{\'{e}}nez-Gal{\'{a}}n, Zhavoronkov, Schloz,
  Morales, and Ivanov}}]{Jimenez2017}
\bibinfo{author}{\bibfnamefont{{\'{A}}.}~\bibnamefont{Jim{\'{e}}nez-Gal{\'{a}}n}},
  \bibinfo{author}{\bibfnamefont{N.}~\bibnamefont{Zhavoronkov}},
  \bibinfo{author}{\bibfnamefont{M.}~\bibnamefont{Schloz}},
  \bibinfo{author}{\bibfnamefont{F.}~\bibnamefont{Morales}}, \bibnamefont{and}
  \bibinfo{author}{\bibfnamefont{M.}~\bibnamefont{Ivanov}},
  \bibinfo{journal}{Optics Express} \textbf{\bibinfo{volume}{25}},
  \bibinfo{pages}{22880} (\bibinfo{year}{2017}),
  \urlprefix\url{http://www.opticsexpress.org/abstract.cfm?URI=oe-25-19-22880}.

\bibitem[{\citenamefont{Floss et~al.}(2018)\citenamefont{Floss, Lemell,
  Wachter, Smejkal, Sato, Tong, Yabana, and Burgd{\"{o}}rfer}}]{Floss2018}
\bibinfo{author}{\bibfnamefont{I.}~\bibnamefont{Floss}},
  \bibinfo{author}{\bibfnamefont{C.}~\bibnamefont{Lemell}},
  \bibinfo{author}{\bibfnamefont{G.}~\bibnamefont{Wachter}},
  \bibinfo{author}{\bibfnamefont{V.}~\bibnamefont{Smejkal}},
  \bibinfo{author}{\bibfnamefont{S.~A.} \bibnamefont{Sato}},
  \bibinfo{author}{\bibfnamefont{X.-M.} \bibnamefont{Tong}},
  \bibinfo{author}{\bibfnamefont{K.}~\bibnamefont{Yabana}}, \bibnamefont{and}
  \bibinfo{author}{\bibfnamefont{J.}~\bibnamefont{Burgd{\"{o}}rfer}},
  \bibinfo{journal}{Phys. Rev. A} \textbf{\bibinfo{volume}{97}},
  \bibinfo{pages}{11401} (\bibinfo{year}{2018}),
  \urlprefix\url{https://link.aps.org/doi/10.1103/PhysRevA.97.011401}.

\bibitem[{\citenamefont{Vampa et~al.}(2014)\citenamefont{Vampa, McDonald,
  Orlando, Klug, Corkum, and Brabec}}]{Vampa2014}
\bibinfo{author}{\bibfnamefont{G.}~\bibnamefont{Vampa}},
  \bibinfo{author}{\bibfnamefont{C.}~\bibnamefont{McDonald}},
  \bibinfo{author}{\bibfnamefont{G.}~\bibnamefont{Orlando}},
  \bibinfo{author}{\bibfnamefont{D.}~\bibnamefont{Klug}},
  \bibinfo{author}{\bibfnamefont{P.}~\bibnamefont{Corkum}}, \bibnamefont{and}
  \bibinfo{author}{\bibfnamefont{T.}~\bibnamefont{Brabec}},
  \bibinfo{journal}{Physical Review Letters} \textbf{\bibinfo{volume}{113}},
  \bibinfo{pages}{73901} (\bibinfo{year}{2014}),
  \urlprefix\url{https://link.aps.org/doi/10.1103/PhysRevLett.113.073901}.

\bibitem[{\citenamefont{Huang et~al.}(2018)\citenamefont{Huang,
  Hern{\'a}ndez-Garc{\'\i}a, Huang, Huang, Lu, Rego, Hickstein, Ellis,
  Jaron-Becker, Becker et~al.}}]{Huang:2018aa}
\bibinfo{author}{\bibfnamefont{P.-C.} \bibnamefont{Huang}},
  \bibinfo{author}{\bibfnamefont{C.}~\bibnamefont{Hern{\'a}ndez-Garc{\'\i}a}},
  \bibinfo{author}{\bibfnamefont{J.-T.} \bibnamefont{Huang}},
  \bibinfo{author}{\bibfnamefont{P.-Y.} \bibnamefont{Huang}},
  \bibinfo{author}{\bibfnamefont{C.-H.} \bibnamefont{Lu}},
  \bibinfo{author}{\bibfnamefont{L.}~\bibnamefont{Rego}},
  \bibinfo{author}{\bibfnamefont{D.~D.} \bibnamefont{Hickstein}},
  \bibinfo{author}{\bibfnamefont{J.~L.} \bibnamefont{Ellis}},
  \bibinfo{author}{\bibfnamefont{A.}~\bibnamefont{Jaron-Becker}},
  \bibinfo{author}{\bibfnamefont{A.}~\bibnamefont{Becker}},
  \bibnamefont{et~al.}, \bibinfo{journal}{Nature Photonics}
  \textbf{\bibinfo{volume}{12}}, \bibinfo{pages}{349} (\bibinfo{year}{2018}),
  \urlprefix\url{https://doi.org/10.1038/s41566-018-0145-0}.

\bibitem[{\citenamefont{Pisanty and
  Jim{\'e}nez-Gal{\'a}n}(2017)}]{Pisanty:2017aa}
\bibinfo{author}{\bibfnamefont{E.}~\bibnamefont{Pisanty}} \bibnamefont{and}
  \bibinfo{author}{\bibfnamefont{{\'A}.}~\bibnamefont{Jim{\'e}nez-Gal{\'a}n}},
  \bibinfo{journal}{Physical Review A} \textbf{\bibinfo{volume}{96}},
  \bibinfo{pages}{063401} (\bibinfo{year}{2017}),
  \urlprefix\url{https://link.aps.org/doi/10.1103/PhysRevA.96.063401}.

\bibitem[{\citenamefont{Jim{\'{e}}nez-Gal{\'{a}}n
  et~al.}(2018)\citenamefont{Jim{\'{e}}nez-Gal{\'{a}}n, Dixit, Patchkovskii,
  Smirnova, Morales, and Ivanov}}]{Jimenez-Galan2018}
\bibinfo{author}{\bibfnamefont{{\'{A}}.}~\bibnamefont{Jim{\'{e}}nez-Gal{\'{a}}n}},
  \bibinfo{author}{\bibfnamefont{G.}~\bibnamefont{Dixit}},
  \bibinfo{author}{\bibfnamefont{S.}~\bibnamefont{Patchkovskii}},
  \bibinfo{author}{\bibfnamefont{O.}~\bibnamefont{Smirnova}},
  \bibinfo{author}{\bibfnamefont{F.}~\bibnamefont{Morales}}, \bibnamefont{and}
  \bibinfo{author}{\bibfnamefont{M.}~\bibnamefont{Ivanov}},
  \bibinfo{journal}{Nature Communications} \textbf{\bibinfo{volume}{9}},
  \bibinfo{pages}{850} (\bibinfo{year}{2018}), ISSN \bibinfo{issn}{2041-1723},
  \urlprefix\url{https://doi.org/10.1038/s41467-018-03167-2}.

\bibitem[{\citenamefont{Hern\'andez-Garc\'{\i}a
  et~al.}(2016)\citenamefont{Hern\'andez-Garc\'{\i}a, Durfee, Hickstein,
  Popmintchev, Meier, Murnane, Kapteyn, Sola, Jaron-Becker, and
  Becker}}]{Hernandez-Garcia2016}
\bibinfo{author}{\bibfnamefont{C.}~\bibnamefont{Hern\'andez-Garc\'{\i}a}},
  \bibinfo{author}{\bibfnamefont{C.~G.} \bibnamefont{Durfee}},
  \bibinfo{author}{\bibfnamefont{D.~D.} \bibnamefont{Hickstein}},
  \bibinfo{author}{\bibfnamefont{T.}~\bibnamefont{Popmintchev}},
  \bibinfo{author}{\bibfnamefont{A.}~\bibnamefont{Meier}},
  \bibinfo{author}{\bibfnamefont{M.~M.} \bibnamefont{Murnane}},
  \bibinfo{author}{\bibfnamefont{H.~C.} \bibnamefont{Kapteyn}},
  \bibinfo{author}{\bibfnamefont{I.~J.} \bibnamefont{Sola}},
  \bibinfo{author}{\bibfnamefont{A.}~\bibnamefont{Jaron-Becker}},
  \bibnamefont{and} \bibinfo{author}{\bibfnamefont{A.}~\bibnamefont{Becker}},
  \bibinfo{journal}{Phys. Rev. A} \textbf{\bibinfo{volume}{93}},
  \bibinfo{pages}{043855} (\bibinfo{year}{2016}),
  \urlprefix\url{https://link.aps.org/doi/10.1103/PhysRevA.93.043855}.

\bibitem[{\citenamefont{Bandrauk et~al.}(2017)\citenamefont{Bandrauk, Guo, and
  Yuan}}]{Bandrauk_2017}
\bibinfo{author}{\bibfnamefont{A.~D.} \bibnamefont{Bandrauk}},
  \bibinfo{author}{\bibfnamefont{J.}~\bibnamefont{Guo}}, \bibnamefont{and}
  \bibinfo{author}{\bibfnamefont{K.-J.} \bibnamefont{Yuan}},
  \bibinfo{journal}{Journal of Optics} \textbf{\bibinfo{volume}{19}},
  \bibinfo{pages}{124016} (\bibinfo{year}{2017}),
  \urlprefix\url{https://doi.org/10.1088{\%}2F2040-8986{\%}2Faa9673}.

\end{thebibliography}

\end{document}